\documentclass[A4paper]{article}
\usepackage{amsmath}
\usepackage{amssymb}
\usepackage{graphicx,psfrag,epsf}
\usepackage{enumerate}
\usepackage{natbib}
\usepackage{url} 
\usepackage{fullpage}

\newcommand{\blind}{1}

\renewcommand\footnotemark{}

\newcommand{\bbeta}{\boldsymbol{\beta}}
\newcommand{\bdelta}{\boldsymbol{\delta}}
\newcommand{\bgamma}{\boldsymbol{\gamma}}
\newcommand{\bomega}{\boldsymbol{\omega}}

\newcommand{\bpsi}{\boldsymbol{\psi}}
\newcommand{\btheta}{\boldsymbol{\theta}}

\newcommand{\ba}{\mathbf{a}}
\newcommand{\bA}{\mathbf{A}}
\newcommand{\bd}{\mathbf{d}}
\newcommand{\bD}{\mathbf{D}}

\newcommand{\bX}{\mathbf{X}}
\newcommand{\by}{\mathbf{y}}
\newcommand{\bz}{\mathbf{z}}

\newcommand{\bzero}{\mathbf{0}}

\newcommand{\mY}{\mathcal{Y}}
\newcommand{\mD}{\mathcal{D}}

\newcommand{\E}{\mathrm{E}}
\newcommand{\N}{\mathrm{N}}
\newcommand{\T}{\mathrm{T}}

\newcommand{\given}{\,|\,}

\newcommand{\mi}[1]{\mathrm{\textit{#1}}}

\DeclareMathOperator*{\argmax}{arg\,max}
\DeclareMathOperator*{\I}{I}

\begin{document}

\def\spacingset#1{\renewcommand{\baselinestretch}%
{#1}\small\normalsize} \spacingset{1}


\if1\blind
{
  \title{\bf \large Bayesian Design of Experiments using\\ Approximate Coordinate Exchange
  \vspace{-.5cm}
  }
  \author{\normalsize Antony M. Overstall\hspace{.2cm}\\
    \normalsize School of Mathematics and Statistics,\\ \normalsize University of Glasgow,\\ \normalsize Glasgow G12 8QW UK\\ \normalsize (Antony.Overstall@glasgow.ac.uk) \\[1ex]
    \normalsize David C. Woods \\
    \normalsize Southampton Statistical Sciences Research Institute,\\ \normalsize University of Southampton,\\ \normalsize Southampton SO17 1BJ UK \\ \normalsize (D.Woods@southampton.ac.uk)}
    \date{\vspace{-1.2cm}}
  \maketitle
} \fi

\if0\blind
{
  \bigskip
  \bigskip
  \bigskip
  \begin{center}
    {\LARGE\bf Title}
\end{center}
  \medskip
} \fi

\bigskip
\begin{abstract}
The construction of decision-theoretic Bayesian designs for realistically-complex nonlinear models is computationally challenging, as it requires the optimization of analytically intractable expected utility functions over high-dimensional design spaces. We provide the most general solution to date for this problem through a novel approximate coordinate exchange algorithm. This methodology uses a Gaussian process emulator to approximate the expected utility as a function of a single design coordinate in a series of conditional optimization steps. It has flexibility to address problems for any choice of utility function and for a wide range of statistical models with different numbers of variables, numbers of runs and randomization restrictions. In contrast to existing approaches to Bayesian design, the method can find multi-variable designs in large numbers of runs without resorting to asymptotic approximations to the posterior distribution or expected utility. The methodology is demonstrated on a variety of challenging examples of practical importance, including design for pharmacokinetic models and design for mixed models with discrete data. For many of these models, Bayesian designs are not currently available. Comparisons are made to results from the literature, and to designs obtained from asymptotic approximations.

\end{abstract}

\noindent%
\vfill
\hfill {\tiny technometrics tex template (do not remove)}

\newpage
\section[Introduction]{Introduction}
\label{INTRO}


Bayesian design of experiments is a natural paradigm for many problems arising in the physical sciences and engineering, particularly those concerning the estimation of nonlinear models where design performance, as measured by classical optimality criteria, is dependent on the a priori unknown values of the model parameters. A decision-theoretic approach, reviewed by \citet{CV1995}, determines an optimal allocation of experimental resources via maximization of the expected utility
\begin{equation}\label{eq:exputil}
U(\bdelta) = \int\int_{\Psi,\mY} u(\bdelta,\bpsi,\by)\pi(\by,\bpsi\given\bdelta)\,\mathrm{d}\by\mathrm{d}\bpsi\,.
\end{equation}
Here, the utility $u(\bdelta,\bpsi,\by)$ quantifies the experimenter's gain from using design $\boldsymbol{\delta}\in\mD$ to obtain data $\by\in\mY$ assuming model parameter values $\bpsi\in\Psi$, with the statistical model defined through the joint density function $\pi(\by,\bpsi\given\bdelta) = \pi(\by\given\bpsi,\bdelta)\pi(\bpsi)$. As an example, assume the $i$th response is modeled as $y_i = g(x_i;\,\btheta) + \varepsilon_i$ with the $x_i$ defining values taken by a controllable variable, $\btheta$ being a vector of unknown parameters defining the mean response, and observation error $\varepsilon_i\sim N(0,\sigma^2)$ $(i=1,\ldots,n)$. Then $\bpsi^{\T} = (\btheta^{\T}, \sigma^2)$, $\bdelta = (x_1,\ldots,x_n)^{\T}$ and likelihood $\pi(\by\given \bpsi,\bdelta)$ is a multivariate normal density function. The utility function $u(\bdelta,\bpsi,\by)$ will typically be a function of some posterior quantities of $\bpsi$ (see Section~\ref{UTILFUNC}).   

Selection of a \textit{fully-Bayesian} optimal design $\bdelta^\star = \mbox{arg max}_{\bdelta\in\mD}U(\bdelta)$ has traditionally been challenging for all but the most straightforward utility functions and models due to the high-dimensional and, typically, analytically intractable integrals in~\eqref{eq:exputil}. Some recent progress has been made using simulation-based methodologies for low-dimensional problems, i.e. small numbers of controllable variables and/or small numbers of design points, see \citet{RDMP2015} and references therein. There are, however, no methods available for decision-theoretic Bayesian optimal, or near-optimal, multi-variable design for nonlinear models. The methodology in this paper fills this important gap, and is demonstrated on generic problems of practical importance including pharmacokinetic studies and experiments that produce discrete data. Previous attempts to obtain fully-Bayesian optimal designs for these types of experiment have been extremely limited.

In a landmark paper for low-dimensional design problems, \citet{MP1996} proposed selection of a design by maximizing a surrogate function found by approximating $U(\bdelta)$ for a small number, $m$, of designs using simulation, and then smoothing the resulting values, $\tilde{U}(\bdelta_1),\ldots,\tilde{U}(\bdelta_m$). See also \citet{JGJR2016} and \citet{WWAH2016}. In essence, these approaches perform a computer experiment to construct a statistical \textit{emulator} for the approximation $\tilde{U}(\bdelta)$, a research area where there has been huge activity in recent years (see, for example, \citealp{DMSB2015}, Section~V). For an experiment with $n$ runs and $v$ variables, $\bdelta$ has $nv$ elements. Therefore, application of this approach to design for multi-variable models suffers from a curse of dimensionality, requiring (i) the construction of emulators in very high dimensions; (ii) large, e.g. space-filling, designs composed of selections of points from an $nv$-dimensional space, leading to (iii) a prohibitive number of evaluations of $\tilde{U}(\bdelta)$, particularly if $\tilde{U}(\bdelta)$ is computationally expensive. 

Our approach overcomes these problems by building a series of one-dimensional emulators for the approximated expected utility. We emulate $\tilde{U}(\bdelta) = \tilde{U}(\delta_i\given\bdelta_{(i)})$  as a function of only the $i$th ``coordinate'' (or element) $\delta_i$ conditional on $\bdelta_{(i)} = (\delta_1, \ldots, \delta_{i-1}, \delta_{i+1}, \delta_{nv})^{\T}$, the values of all coordinates excluding the $i$th $(i = 1,\ldots,nv)$. When these emulators are combined with a continuous version of the coordinate exchange algorithm \citep{MN1995}, an effective and computationally efficient design selection methodology results. Conditional, coordinate-wise, optimization is key to overcoming the curse of dimensionality described above.

Until relatively recently, the usual approach to Bayesian design was to use a normal distribution as an asymptotic approximation to the posterior distribution of $\boldsymbol{\psi}$ (e.g. \citealp{CL1989}). For standard utility functions (see Section~\ref{normalapprox}), use of such a \textit{pseudo-Bayesian} approach leads to the integrand in~\eqref{eq:exputil} no longer depending on the data $\by$. The resulting integral, with respect to $\bpsi$, typically has much lower dimension and can be approximated using efficient deterministic quadrature rules \citep{GJS2009}. However, the appropriateness of such approximations for small experiments is open to question.

For high-dimensional design, an alternative to the use of a normal approximation was suggested by \citet{R2014}. These authors combined the simulation-based approach of \citet{M1999} (see also \citealp{MSI2004} and \citealp{ABPR2006}) with a dimension-reduction scheme to find designs for single-variable nonlinear models ($v=1$) variables and a large number of runs. Designs were restricted to those formed from a sampling scheme defined via two parameters; for example, the initial design point and a spacing parameter. An optimal design in this subclass then consists of the best choices of these two parameters, a substantially easier optimisation problem to solve. 

In contrast to either applying an asymptotic approximation or restricting attention to a subset of the design space, both of which may result in the selection of inefficient designs with respect to the exact expected utility, we attempt to find optimal or efficient designs for the original problem across the whole design space via an approximate optimization scheme. These three different approaches are compared in Section~\ref{EXAMPLES}.

The remainder of the paper is organised as follows. In Section~\ref{ACE} we describe approximate coordinate exchange for finding decision-theoretic Bayesian designs, including the use of Monte Carlo integration and Gaussian process emulators to approximate the expected utility. The methods are applied to a range of challenging and practically relevant examples in Section~\ref{EXAMPLES} including models for which Bayesian design has previously been computationally infeasible. We summarise the advantages of our approach in Section~\ref{DISC} and highlight some ongoing work.

\section[ACE Algorithm]{Approximate Coordinate Exchange (ACE)}
\label{ACE}

We first establish some notation. Suppose that a design consists of $n$ runs or points, each of which determines the settings of $v$ controllable variables and results in a single observation of the response variable. Let $\bD$ denote the $n \times v$ design matrix with $k$th row $\bd_k$ specifying the settings of the $v$ factors in the $k$th run $(k=1,\ldots,n)$. Let $q=nv$, then the design may be represented as a $q$-vector $\bdelta = \mathrm{vec}\left(\bD\right) \in \mD\subset\mathbb{R}^{q}$, where $\mathrm{vec}(\cdot)$ denotes vectorisation via stacking the columns of a matrix and $\mD$ is the $q$-dimensional design space. 

The proposed algorithm for decision-theoretic Bayesian design has two phases. Phase I applies a novel coordinate exchange algorithm where, for each coordinate, maximisation of $U(\bdelta)$ is replaced by maximisation of a surrogate function $\hat{U}(\bdelta)$. Phase I tends to produce clusters of design points that are very similar in the values of the controllable variables. Such clustering is common in heuristic design search \citep[see also][]{GJS2009}. Hence, in Phase II, we check if the points in each cluster can be consolidated into a replicated design point using a point exchange algorithm \citep[][ch.~12]{ADT2007}. Replication of points is common in optimal design for parametric models and a key principle of design of experiments (\citealp{WH2009}, ch.~1). In Phase II, the candidate set is the design found from Phase I. The two phases form an approximate coordinate and point exchange algorithm which, for brevity, we call the ACE algorithm.

In Section~\ref{ACEalg} we define the ACE algorithm. For Steps~\ref{designstep}~-~\ref{emstep} of the algorithm, we assume the availability of (i) a Monte Carlo approximation of the expected utility, 
$$
\tilde{U}(\delta_i\given\bdelta_{(i)}) = \tilde{U}(\bdelta) = \sum_{l = 1}^B u(\bdelta,\by_l,\bpsi_l)/B\,,
$$ 
with $\{\by_l,\bpsi_l\}_{l=1}^B$ a random sample from the joint distribution with density $\pi(\by,\bpsi\given\bdelta)$; (ii) \textit{coordinate-designs} $\xi_i = \{\delta_i^1,\ldots,\delta_i^m\}\in\mD_i$ at which we evaluate $\tilde{U}(\delta_i\given\bdelta_{(i)})$, where $\mD_i\subset\mathbb{R}$ is the domain for the $i$th coordinate; and (iii) a suitable one-dimensional emulator, $\hat{U}(\delta_i \given \bdelta_{(i)})$, for $\tilde{U}(\delta_i \given \bdelta_{(i)})$. Further details are given in Section~\ref{GPE}, with examples in Section~\ref{TOYEX}.

ACE is designed to solve a stochastic optimization problem, as only approximations to the expected utility are available formed as linear combinations of realisations of the random variable $u(\bdelta,\by,\bpsi)$. As such, proposed changes to the design in Steps~\ref{acceptstep1} and~\ref{acceptstep2} of the algorithm are accepted with probability derived from a Bayesian test of the difference in the means of Monte Carlo approximations to the expected utility for the current and proposed designs. Further details are given in Section~\ref{acceptance}.


\subsection{The ACE algorithm}\label{ACEalg}

\begin{enumerate}
\setcounter{enumi}{-1}
\item
Choose an initial design $\bdelta^0$ and set the current design $\bdelta^C = \bdelta^0$.
\begin{center}
\textit{Phase I: coordinate exchange}
\end{center}
\item
For $i=1,\dots,q$, \label{mainstep}
				\begin{enumerate}
				\item\label{designstep} Select an $m$ point coordinate-design $\xi_i = \{\delta_i^1,\ldots,\delta_i^m\}\in\mD_i$.
				\item\label{evalstep} Evaluate $\tilde{U}(\delta_i^1 \given \bdelta^C_{(i)}),\ldots,\tilde{U}(\delta_i^m \given \bdelta^C_{(i)})$.
				\item\label{emstep} Construct $\hat{U}(\delta_i \given \bdelta^C_{(i)})$ by fitting a statistical model to $\left\{\delta_i^j, \tilde{U}(\delta_i^j)\right\}_{j=1}^m$.
				\item\label{acceptstep1} With probability $p^\dagger_{\mathrm{\textit{I}}}$, set $\delta^C_i = \delta^\dagger_i = \argmax_{\delta_i\in\mD_i}\hat{U}(\delta_i \given \bdelta^C_{(i)})$, where
				\begin{equation}\label{p1}
				p^\dagger_{\mathrm{\textit{I}}} = 1 - T_{2B - 2}\left(-\frac{B\tilde{U}(\bdelta^{C\dagger}) - B\tilde{U}(\bdelta^C)}{\sqrt{2B \hat{\nu}_{\mathrm{\textit{I}}}}}\right)\,,
				\end{equation}
				$T_{2B-2}$ is the probability distribution function for a $t$ distribution with $2B-2$ degrees of freedom, $\bdelta^{C\dagger} = (\delta^C_1,\ldots,\delta^C_{i-1}, \delta_i^\dagger,\delta^C_{i+1},\ldots,\delta^C_q)^{\T}$, and
				$$
				\hat{\nu}_{\mathrm{\textit{I}}} = \frac{\sum_{l=1}^B \left[u(\bdelta^{C\dagger}, \by^\dagger_l, \bpsi^\dagger_l) - \tilde{U}(\bdelta^{C\dagger})\right]^2 + \sum_{l=1}^B \left[u(\bdelta^C, \by^C_l, \bpsi^C_l) - \tilde{U}(\bdelta^C)\right]^2}{2B - 2}\,,
				$$
				for $\{\by^\dagger_l, \bpsi_l^\dagger\}_{l=1}^B$ and $\{\by^C_l, \bpsi_l^C\}_{l=1}^B$ independent random samples from $\pi(\by,\bpsi\given\bdelta^{C\dagger})$ and $\pi(\by,\bpsi\given\bdelta^C)$, respectively.
				\end{enumerate}
\item
Repeat step~\ref{mainstep} $N_{\mathrm{\textit{I}}}$ times. 
\begin{center}
\textit{Phase II (point exchange)}
\end{center}
\item\label{PEstep}
			\begin{enumerate}
			\item
			For $k=1,\dots,n$, let $\boldsymbol{\delta}_k^{(1)} = \mathrm{vec}(\mathbf{D}^{(1)}_k)$, where 
			$$\bD^{(1)}_k = \left[\left(\bD^{C}\right)^\T\,\,\left(\bd^C_k\right)^\T\right]^\T\,,$$
			and $\bd^C_k$ is the $k$th row of $\bD^C$, the design matrix for $\bdelta^C$.
			
			\item
			Let $k^\dagger = \argmax_k \tilde{U}(\boldsymbol{\delta}_k^{(1)})$ and set $\bD^{{(2)}} = \mathbf{D}_{k^\dagger}^{(1)}$.
			
			\item
			For $h=1,\dots,n+1$, let $\boldsymbol{\delta}_s^{{(3)}} = \mathrm{vec}(\mathbf{D}^{{(3)}}_h)$, where 
			$$
			\bD^{(3)}_h = \left[\left(\bd^{(2)}_1\right)^\T\,\ldots\,\left(\bd^{(2)}_{h-1}\right)^\T\,\,\left(\bd^{(2)}_{h+1}\right)^\T\,\ldots\,\left(\bd^{(2)}_{n+1}\right)^\T\right]^\T\,,
			$$
			and $\bd^{(2)}_h$ is the $h$th row of $\bD^{(2)}$.
 			\item
 			Let $h^\dagger = \argmax_h \tilde{U}(\boldsymbol{\delta}_h^{{(3)}})$
			
			\item\label{acceptstep2} With probability $p^\dagger_{\mathrm{\textit{II}}}$, set $\bdelta^C = \bdelta^{(3)}_{h^\dagger}$, where
				\begin{equation}\label{p2}
				p^\dagger_{\mathrm{\textit{II}}} = 1 - T_{2B - 2}\left(-\frac{B\tilde{U}(\bdelta^{(3)}_{h^\dagger}) - B\tilde{U}(\bdelta^C)}{\sqrt{2B \hat{\nu}_{\mathrm{\textit{II}}}}}\right)
				\end{equation}
				and
				$$
				\hat{\nu}_{\mathrm{\textit{II}}} = \frac{\sum_{l=1}^B \left[u(\bdelta^{(3)}_{h^\dagger}, \by^{(3)}_l, \bpsi^{(3)}_l) - \tilde{U}(\bdelta^{(3)}_{h^\dagger})\right]^2 + \sum_{l=1}^B \left[u(\bdelta^C, \by^C_l, \bpsi^C_l) - \tilde{U}(\bdelta^C)\right]^2}{2B - 2}\,,
				$$
				with $\{\by^{(3)}_l, \bpsi_l^{(3)}\}_{l=1}^B$ a random sample from $\pi(\by,\bpsi\given\bdelta^{(3)}_{h^\star})$.
			\end{enumerate}
\item Repeat step~\ref{PEstep} $N_{\mathrm{\textit{II}}}$ times.
\item Return $\bdelta^\star = \bdelta^C$.
\end{enumerate}

The decision on when to terminate a run of the algorithm, i.e. choice of $N_{\mathrm{\textit{I}}}$ and $N_{\mathrm{\textit{II}}}$, is complicated by the stochastic nature of the approximation to the expected utility. For the examples in Section~\ref{EXAMPLES}, $N_{\mi{I}} = 20$ and $N_{\mi{II}}=100$ are sufficient to achieve approximate convergence. Here, convergence is assessed graphically from trace plots of $\tilde{U}(\bdelta^C)$ against iteration number; see Section~\ref{COMP_SEC} for examples of such plots. 

To avoid local optima, the algorithm is run $M$ times (in embarrassingly parallel fashion) with each run starting from a different, randomly-chosen, initial design $\bdelta_0$ (a random Latin hypercube design, unless otherwise stated). The selected design, $\bdelta^\star$, is the design having the highest average approximate expected utility, averaged across $C$ sets of Monte Carlo simulations. In this paper, $M=C=20$ was used, unless otherwise stated.  



\subsection{Emulation via computer experiments (steps~\ref{designstep} - \ref{emstep})}
\label{GPE}

In Phase~I of the algorithm, a sequence of one-dimensional emulators is constructed for $\tilde{U}(\delta_i\given\bdelta_{(i)})$, $i = 1,\ldots,q$ (Step~\ref{emstep}). A variety of smoothing or interpolation techniques could be applied to construct each emulator. \citet{MP1996} used local polynomial regression to emulate low-dimensional design utilities. We adopt a Gaussian Process (GP) regression model \citep[see, for example,][]{RW2006}, which is widely used for computer experiments, and use the posterior predictive mean as an emulator. Let 
$$
\hat{\mu}_i = \sum_{j=1}^m\tilde{U}(\delta_i^j\given\bdelta^C_{(i)})/m\,,
$$
$$
\hat{\sigma}^2_i = \sum_{j=1}^m\left(\tilde{U}(\delta_i^j\given\bdelta^C_{(i)}) - \hat{\mu}_i\right)^2/(m-1)\,,
$$ 
and $z(\delta_i) = \left(U(\delta_i \given \bdelta^C_{(i)}) - \hat{\mu}_i\right)/\hat{\sigma}_i$ for any $\delta_i\in\mD_i$. The GP model assumes that any vector $\bz(\zeta) = \left[z(\delta^1),\ldots,z(\delta^{m_0})\right]^\T$, for $\zeta = \{\delta^1,\ldots,\delta^{m_0}\}\in\mD_i^{m_0}$ and integer $m_0$, has joint distribution $\mathrm{N}\left(\bzero_{m_0}, \bA(\zeta)\right)$, with $\bzero_{m_0}$ the $m_0$ zero-vector and $\bA(\zeta)$ an $m_0\times m_0$ covariance matrix.  Hence, the posterior predictive mean of $\tilde{U}(\delta\given\bdelta_{(i)})$ at an arbitrary $\delta\in\mD_i$ can be derived using standard results on the conditional distribution of normal random variables and used as an emulator: 
\begin{equation*}
\begin{split}
\hat{U}\left(\delta \given \bdelta_{(i)}^C\right) & = \hat{\mu}_i + \hat{\sigma}_i\E\left[z(\delta) \given \bz(\xi_i)\right]\\
& = \hat{\mu}_i + \hat{\sigma}_i \ba(\delta, \xi_i)^\T \mathbf{A}(\xi_i)^{-1}\bz(\xi_i)\,. 
\end{split}
\end{equation*}
Under the common assumption of a squared exponential correlation function, $\bA(\xi_i)$ and $\ba(\delta,\xi_i)$ have entries
\begin{equation*}
\begin{split}
\bA(\xi_i)_{st} & = \exp\left\{-\rho(\delta_i^s-\delta_i^t)^2\right\} + \eta\I(r=s)\,, \\
\ba(\delta,\xi_i)_s & = \exp\left\{-\rho(\delta_i^s-\delta)^2\right\}\,,
\end{split}
\end{equation*}
for $s,t = 1,\ldots, m$, where $\I(E)$ is the indicator function for the event $E$, and $\rho, \eta>0$ are unknown parameters. The inclusion of nugget $\eta$ ensures the emulator will smooth, rather than interpolate, the Monte Carlo approximations of the expected utility. To limit computational complexity, at each iteration we find maximum likelihood estimates of $\rho$ and $\eta$ via Fisher scoring \citep[see, for example, ][pp. 174-177]{pawitan}. In contrast, a fully-Bayesian approach would require application of a Markov chain Monte Carlo algorithm to construct each emulator, substantially increasing the computational cost of the algorithm.

At each iteration of Step~\ref{designstep}, a coordinate-design $\xi_i = (\delta_i^1,\ldots,\delta_i^m)$ must be chosen at which to evaluate $\tilde{U}(\delta_i\given\bdelta_{(i)})$. We use a space-filling design, specifically a randomly-selected one-dimensional Latin hypercube design \citep[see, for example, ][ch.~5]{SWN2003}, constructed by dividing $\mD_i$ into $m$ equally-sized sub-intervals, and then generating a point at random from each interval. We set $m=20$, unless otherwise stated. This choice of $m$ is conservative relative to the rule of thumb \citep{LSW2009} of setting $m$ equal to 10 times the number of input dimensions (suggesting $m=10$ in our case). We have, however, found it works well in practice for a variety of different types of examples, giving accurate emulators and not being overly computationally demanding. 

\subsection{Adjusting a design coordinate (step~\ref{acceptstep1}) or point (step~\ref{acceptstep2})}\label{acceptance}
To make a change to the $i$th coordinate in Step~\ref{acceptstep1}, we first find $\delta_i^\dagger$, the value of the coordinate that maximizes the emulator. We find the maximum by evaluating $\hat{U}\left(\delta \given \bdelta_{(i)}^C\right)$ for 10,000 uniformly generated points in $\mD_i$. This discretization of the problem has proved both more reliable than continuous optimization and sufficiently computationally efficient.    

Choice of $\delta_i^\dagger$ is subject to both Monte Carlo error, from the evaluation of $\tilde{U}(\delta_i\given\bdelta_{(i)})$, and emulator error from the estimation of $\hat{U}(\delta_i\given\bdelta_{(i)})$ resulting, for example, from an inappropriate choice of correlation function or errors in estimating $\rho$ and $\eta$. It is clearly impossible to use usual residual diagnostics \citep{BOH2009} to check emulator adequacy at each iteration of the algorithm. Instead, emulator error is eliminated from the decision to adjust a design coordinate by performing additional Monte Carlo integration to calculate the probability $p^\dagger_{\mathrm{\textit{I}}}$ in~\eqref{p1}. This quantity is the posterior probability that $\E\left[u(\bpsi,\by,\bdelta^{C\dagger})\right] > \E\left[u(\bpsi,\by,\bdelta^C)\right]$ under non-informative prior distributions and using Monte Carlo samples $\left\{\boldsymbol{\psi}_l^C,\mathbf{y}_l^C\right\}_{l=1}^B$ and $\left\{\boldsymbol{\psi}^\dagger_l,\mathbf{y}^\dagger_l\right\}_{l=1}^B$, assuming both $u(\bpsi^\dagger,\by^\dagger,\bdelta^{C\dagger})$ and $u(\bpsi^C,\by^C,\bdelta^C)$ are normally distributed with equal variances. See also \citet{WZ2006} for use of a classical hypothesis test in a simulated annealing algorithm. If this normality assumption were severely violated, a more sophisticated test procedure could be adopted at greater computational cost.

A similar test is performed at Step~\ref{acceptstep2} in Phase 2 of the algorithm to calculate $p^\dagger_{\mathrm{\textit{II}}}$ in~\eqref{p2}. We demonstrate the effect of Step~\ref{acceptstep1} in Section~\ref{poisson}.

\subsection{Illustrative example}
\label{TOYEX}
\label{poisson}

\begin{figure}
\centering
\includegraphics[trim = {0, 5cm, 0, 6cm}, clip, width=140mm]{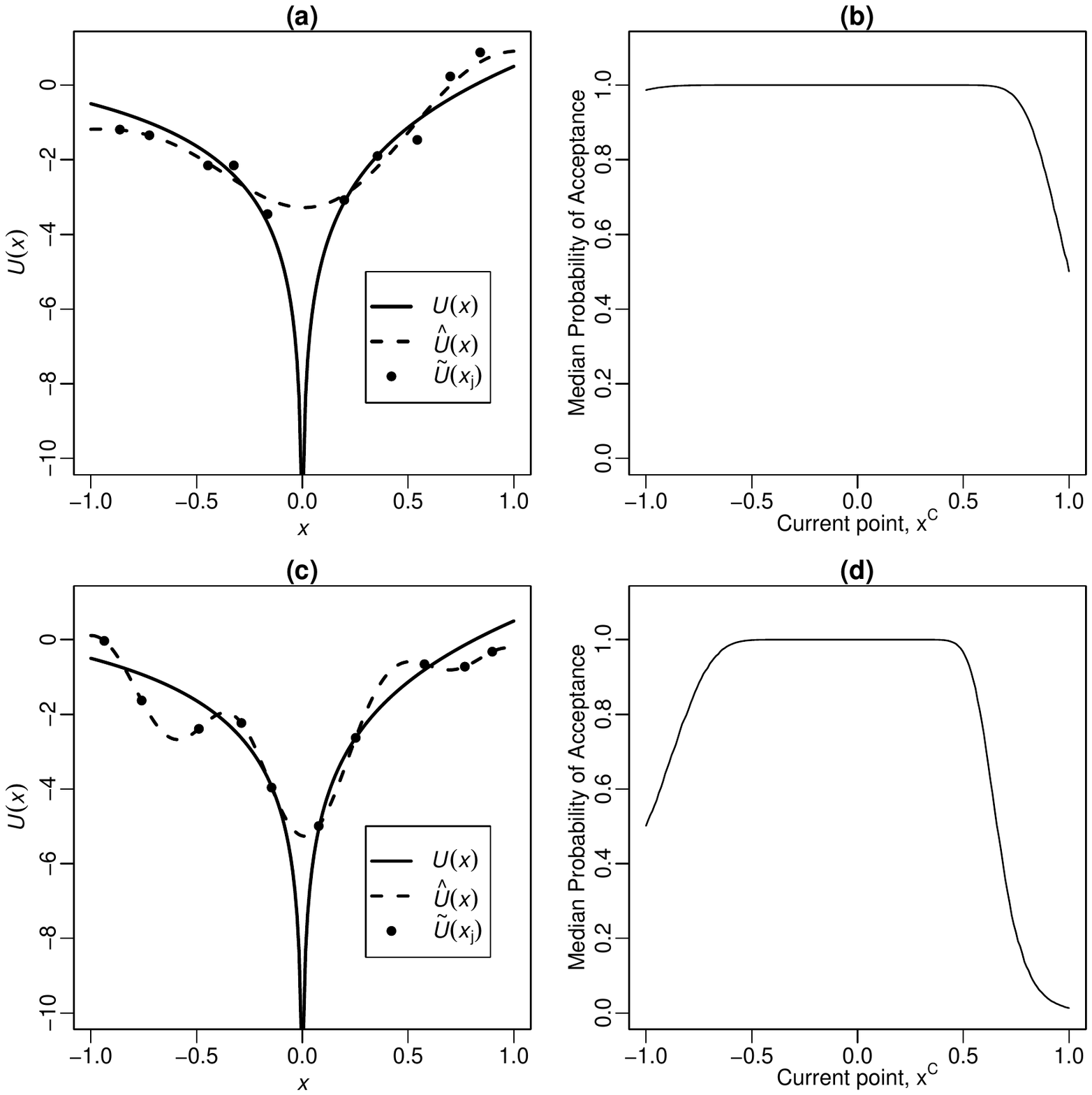}
\caption{\label{TOY_PIC}Poisson example in Section \ref{poisson}. (a), (c) expected utility $U(x)$ against $x$, with Monte Carlo evaluations $\tilde{U}(x)$ at the coordinate-design points and GP emulator $\hat{U}(x)$; (b), (d) median probability $p^\dagger_{\mi{I}}$ of accepting the candidate point against the current point, $x^C$. [Coordinate-designs are: $\xi^1$ for (a), (b); $\xi^2$ for (c), (d)].}
\end{figure}

In this section, we illustrate the ACE methodology, in particular the combination of Steps~\ref{emstep} and~\ref{acceptstep1} in selecting and accepting a proposed change to the design. To enable assessment of the algorithm, we consider the analytically tractable problem of finding a one-point optimal design for the single-variable Poisson model $y|\beta \sim \mathrm{Poisson}(e^{\beta x})$. There is a single design coordinate, $\bdelta = x \in [-1,1]$, and hence our notation is simplified by replacing $\bdelta$ by $x$ in this example. A priori, we assume $\beta\sim\mathrm{N}(0.5, 1)$ and adopt the utility function that leads to pseudo-Bayesian $D$-optimality (Section~\ref{normalapprox}), given by
\begin{eqnarray*}
u(\beta,y,x) & = & \log \mathcal{I}(\beta;\,x)\\
& = & 2\log |x| + \beta x\,,
\end{eqnarray*}
where $\mathcal{I}(\beta;\,x)$ denotes the Fisher information. The expected utility is $U(x) = 2 \log |x| + 0.5 x$ and the optimal design is $x^\star = 1$. 

To simulate one iteration of Phase I of the ACE algorithm, we generate a coordinate-design $\xi^1 = \{x^1,\dots,x^m\}$ as a Latin hypercube and, for each $x^j$, evaluate
$$\tilde{U}(x^j) = 2 \log |x^j| + \frac{x^j}{B}\sum_{i=1}^B\beta_l\,,$$
where $\left\{\beta_l\right\}_{l=1}^B$ for $B=2$ is a sample from a $\mathrm{N}\left(0.5,1\right)$ distribution. Figure~\ref{TOY_PIC}(a) shows $U(x)$ plotted against $x$ with the points $\{x^j,\tilde{U}(x^j)\}_{j=1}^m$ and the GP emulator $\hat{U}(x)$ superimposed (Steps~\ref{designstep},~\ref{evalstep} and~\ref{emstep}). Clearly $\hat{U}(x)$ is maximised at $x^\dagger=1$, and hence this candidate point should be compared to the current point $x^C$ (Step~\ref{acceptstep1}). Figure~\ref{TOY_PIC}(b) shows the median posterior probability, $p^\dagger_{\mi{I}}$, of accepting this candidate point against $x^C$, calculated from repeated calculation of~\eqref{p1} for multiple Monte Carlo samples. This probability is very close to one for nearly all values of $x^C$ except for $x^C\approx x^\dagger$, where the probability reduces to $1/2$. 

For a second coordinate-design, $\xi^2$ (a different Latin hypercube), the results in Figures~\ref{TOY_PIC}(c) and~\ref{TOY_PIC}(d) are obtained. Here, the GP emulator could be viewed as inadequate with the estimate of $\eta$ being too small, resulting in near interpolation of the $\tilde{U}(x^j)$. From Figure~\ref{TOY_PIC}(c), $\hat{U}(x)$ is maximised at $x^\dagger=-1$ and hence this becomes the candidate point. The median posterior acceptance probability, Figure~\ref{TOY_PIC}(d), is now only close to one if $\tilde{U}(x^C)$ is low, i.e. $|x^C|<0.5$. Crucially, $x^\dagger$ will be rejected with high probability if $x^C$ is close to the optimal design; at $x^C=1$, the probability drops to zero.

\section[Examples]{Substantive examples}
\label{EXAMPLES}

The ACE algorithm is now used to find decision-theoretic Bayesian designs for three important cases: a compartmental model, (hierarchical) logistic regression, and dose-response under model-averaging. The designs are found for commonly used utility functions and, where possible, compared to existing results. 

\subsection[Utility functions]{Utility functions}
\label{UTILFUNC}
In this section, we assess and compare designs found using variants on two utility functions, Shannon information gain (SIG) and negative squared error loss (NSEL). In practice, the form of the chosen utility function should be driven by the aims of the experiment and may often incorporate a cost function. We assume throughout that the model parameters can be expressed as $\bpsi^{\T} = \left(\btheta^{\T},\bgamma^{\T}\right)$, with $\btheta$ a $p$-vector of parameters of interest and $\bgamma$ a $(P-p)$-vector of nuisance parameters.


The SIG utility for $\btheta$ is given by 
\begin{align}
u^S(\boldsymbol{\theta},\mathbf{y},\boldsymbol{\delta}) & = \log \pi(\boldsymbol{\theta}\given\mathbf{y},\boldsymbol{\delta}) - \log \pi(\boldsymbol{\theta})\nonumber \\
& = \log \pi(\mathbf{y}\given\boldsymbol{\theta},\boldsymbol{\delta}) - \log \pi(\mathbf{y}\given\boldsymbol{\delta})\,,\label{eq:SIG2}
\end{align}
where~\eqref{eq:SIG2} follows from an application of Bayes theorem and is often more useful for computation. A SIG-optimal design maximizes $U^S(\bdelta) = \E_{\bpsi,\by} \left[u^S(\btheta,\by,\bdelta)\right]$. This is equivalent to maximizing the expected Kullback-Liebler distance between the marginal posterior and prior distributions of $\btheta$, and is also equivalent to minimizing the expected entropy of the posterior distribution for $\btheta$.



The NSEL utility for $\btheta$ is given by 
\begin{equation}\label{eq:NSEL}
u^V(\btheta,\by,\bdelta) = - \sum_{w=1}^p \left[\theta_w - \E(\theta_w|\by,\bdelta)\right]^2\,.
\end{equation}
A NSEL-optimal design maximizes the expected utility $U^V(\bdelta)$, which is equivalent to minimizing the expectation of the trace of the posterior covariance matrix of $\btheta$ with respect to the marginal distribution of $\by$.

\subsubsection{Evaluating the expected utility via numerical approximation} \label{numapprox}

For many statistical models, including most nonlinear models, evaluation of $u^S(\btheta,\by,\bdelta)$ and $u^V(\btheta,\by,\bdelta)$ requires numerical approximation. For given values of $\mathbf{y}$ and $\btheta$, the components of~\eqref{eq:SIG2} can be approximated as
$$
\tilde{\pi}(\mathbf{y}\given\boldsymbol{\theta},\boldsymbol{\delta}) = \frac{1}{\tilde{B}} \sum_{b=1}^{\tilde{B}} \pi(\mathbf{y}\given\boldsymbol{\theta},\tilde{\boldsymbol{\gamma}}_b,\boldsymbol{\delta})\,,\qquad
\tilde{\pi}(\mathbf{y}\given\boldsymbol{\delta}) = \frac{1}{\tilde{B}} \sum_{b=1}^{\tilde{B}} \pi(\mathbf{y}\given\tilde{\boldsymbol{\theta}}_b,\tilde{\boldsymbol{\gamma}}_b,\boldsymbol{\delta})\,,
$$
where $\left\{\tilde{\boldsymbol{\theta}}_b,\tilde{\boldsymbol{\gamma}}_b\right\}_{b=1}^{\tilde{B}}$ is a size $\tilde{B}$ random sample from the prior distribution of $\boldsymbol{\psi}=\left(\boldsymbol{\theta},\boldsymbol{\gamma}\right)$. These quantities can be incorporated into a nested, or double-loop, Monte Carlo approximation of $U^S(\boldsymbol{\delta})$:
\begin{equation*}\label{eq:SIGapprox}
 \tilde{U}^S(\boldsymbol{\delta}) = \frac{1}{B} \sum_{l=1}^B \left[\log \tilde{\pi}(\mathbf{y}_l\given\boldsymbol{\theta}_l,\boldsymbol{\delta}) - \log \tilde{\pi}(\mathbf{y}_l\given\boldsymbol{\delta})\right]\,,
 \end{equation*}
with $\left\{\mathbf{y}_l,\boldsymbol{\theta}_l\right\}_{l=1}^{B}$ a sample from the joint distribution of the response and parameters. Intuitively, the ``inner sample'' of size $\tilde{B}$ is used to approximate the two marginal likelihoods in~\eqref{eq:SIG2}, the first marginal to $\bgamma$ and the second to both $\bgamma$ and $\btheta$, and the ``outer sample'' of size $B$ is then used to approximate the expected utility with respect to the joint distribution of $\mathbf{y}$ and $\btheta$. This approximation is biased for $U^S(\bdelta)$ due to the bias in $\log \tilde{\pi}(\mathbf{y}|\boldsymbol{\theta},\boldsymbol{\delta})$ and $\log \tilde{\pi}(\mathbf{y}|\boldsymbol{\delta})$. However, under regularity conditions satisfied by most models of practical importance (\citealp{S2000}, pp.~80--81), this bias is of order $\tilde{B}^{-1}$ \citep{Ryan2003} and hence asymptotically negligible.
 
For NSEL, $\mathrm{E}(\theta_w\given\mathbf{y},\boldsymbol{\delta})$ in~\eqref{eq:NSEL} can be approximated via importance sampling:
$$\tilde{\mathrm{E}}(\theta_w\given\mathbf{y},\boldsymbol{\delta}) = \frac{\sum_{b=1}^{\tilde{B}} \tilde{\theta}_{lw} \pi(\mathbf{y}\given\tilde{\boldsymbol{\theta}}_b,\tilde{\boldsymbol{\gamma}}_b,\boldsymbol{\delta})}{\sum_{b=1}^{\tilde{B}} \pi(\mathbf{y}\given\tilde{\boldsymbol{\theta}}_b,\tilde{\boldsymbol{\gamma}}_b,\boldsymbol{\delta})}\,,$$
where $\left\{ \tilde{\boldsymbol{\theta}}_b,\tilde{\boldsymbol{\gamma}}_b \right\}_{b=1}^{\tilde{B}}$ is a random sample from the prior distribution of $\boldsymbol{\psi}$, and $\tilde{\theta}_{bw}$ is the $w$th element of $\tilde{\boldsymbol{\theta}}_b$. Hence, the following nested Monte Carlo approximation of the expected utility is obtained:
$$\tilde{U}^V(\boldsymbol{\delta}) = - \frac{1}{B} \sum_{l=1}^B \sum_{w=1}^p \left[\theta_{lw} - \tilde{\mathrm{E}}(\theta_w\given\mathbf{y}_l,\boldsymbol{\delta})\right]^2\,,$$
where $\theta_{lw}$ is the $w$th element of $\boldsymbol{\theta}_l$. Here, the inner sample is used to approximate the posterior expectation, and the outer sample used to approximate the expected utility. Importance sampling has commonly been used to estimate posterior quantities for Bayesian design (see \citealp{RDMP2015} and references therein), although the approximation of the expected utility will again be biased due to bias in $\tilde{\mathrm{E}}(\theta_w\given\mathbf{y},\boldsymbol{\delta})^2$. 

In the examples, we set $\tilde{B}=B = 1000$ for the evaluation of $\tilde{U}(\delta_i\given\bdelta_{(i)}^C)$ in Step~\ref{evalstep} of the ACE algorithm (chosen from practical experience). For the comparisons in Steps~\ref{acceptstep1} and~\ref{acceptstep2}, we set $B=\tilde{B} = 20,000$. 

\subsubsection{Evaluating the expected utility via normal approximation}\label{normalapprox}

The following approximations to $U^S(\bdelta)$ and $U^V(\bdelta)$ are commonly used \citep[][ch.~18]{ADT2007}, justified via a normal approximation to the posterior distribution of $\boldsymbol{\psi}$: 
\begin{eqnarray*}
\phi^S(\boldsymbol{\delta}) & = & \mathrm{E}_{\boldsymbol{\psi}} \left( \log |\mathcal{I}(\boldsymbol{\theta};\boldsymbol{\delta},\bgamma)|\right) = \int_{\Theta}  \log |\mathcal{I}(\boldsymbol{\theta};\boldsymbol{\delta},\bgamma)| \pi(\boldsymbol{\psi}) \mathrm{d}\boldsymbol{\psi},\\
\phi^V(\boldsymbol{\delta}) & = & - \mathrm{E}_{\boldsymbol{\psi}} \left[\mathrm{tr} \left\{\mathcal{I}(\boldsymbol{\theta};\boldsymbol{\delta},\bgamma)^{-1}\right\}\right] = - \int_{\Theta}  \mathrm{tr} \left\{\mathcal{I}(\boldsymbol{\theta};\boldsymbol{\delta},\bgamma)^{-1}\right\} \pi(\boldsymbol{\psi}) \mathrm{d}\boldsymbol{\psi},
\end{eqnarray*}
with $\mathcal{I}(\boldsymbol{\theta};\boldsymbol{\delta},\bgamma)$ the Fisher information matrix for $\btheta$, or an approximation thereof. Designs that maximise $\phi^S$ and $\phi^V$ are sometimes referred to as pseudo-Bayesian $D$- and $A$-optimal designs, respectively. Note that these expressions also result from taking expectations of the utility functions 
$$
u^D(\bpsi,\by,\bdelta) =  \log |\mathcal{I}(\btheta;\boldsymbol{\delta},\bgamma)|\,,\quad u^A(\bpsi,\by,\bdelta) =  - \mathrm{tr} \left\{\mathcal{I}(\btheta;\bdelta,\bgamma)\right\}^{-1}\,,
$$
which do not depend on $\by$. Unbiased Monte Carlo approximations to $\phi^S(\bdelta)$ and $\phi^V(\bdelta)$ can be obtained via sampling from the prior distribution for $\bpsi$:
$$
\hat{\phi}^S(\bdelta) = \frac{1}{B}\sum_{l = 1} ^ B \log |\mathcal{I}(\boldsymbol{\theta}_l;\boldsymbol{\delta},\bgamma_l)|\,, \quad
\hat{\phi}^V(\bdelta) = \frac{1}{B}\sum_{l = 1} ^ B \mathrm{tr} \left\{\mathcal{I}(\boldsymbol{\theta}_l;\boldsymbol{\delta},\bgamma_l)^{-1}\right\}\,. 
$$
For comparison of designs, the $D$-efficiency of design $\bdelta_1$ relative to design $\bdelta_2$ is defined as
\begin{equation}\label{eq:Deff}
\mbox{Eff}_{D}(\bdelta_1, \bdelta_2) = 100 \times \exp\left\{ \left[ \hat{\phi}^S(\bdelta_1) - \hat{\phi}^S(\bdelta_2) \right] / p \right\}\,. 
\end{equation}

\subsection[Compartmental model]{Compartmental model}
\label{COMP_SEC}

Compartmental models are applied in pharmacokinetics to study how materials flow through an organism, and have been used extensively to demonstrate optimal design methodology \citep{A1993, GJS2009}. The archetypal design problem is to choose $n$ sampling times $\bdelta = (t_1, \ldots, t_n)^\T$, in hours, at which to measure the concentration in a subject of a previously administered drug. Here, concentration is modeled as $y_i \sim \N\left(a(\btheta)\mu(\btheta;t_i), \sigma^2 b(\btheta;t_i)\right)\,,$
where $\btheta=\left(\theta_1,\theta_2,\theta_3\right)^{\T}$ are the parameters of interest, $\sigma^2>0$ is a nuisance parameter, $a(\cdot)$ and $b(\cdot;\cdot)$ are application-dependent functions, and $\mu(\btheta;t_i) = \exp\left(-\theta_1 t_i\right) - \exp\left(-\theta_2 t_i\right)$.

%
%


\begin{figure}
\centering
\includegraphics[width=160mm]{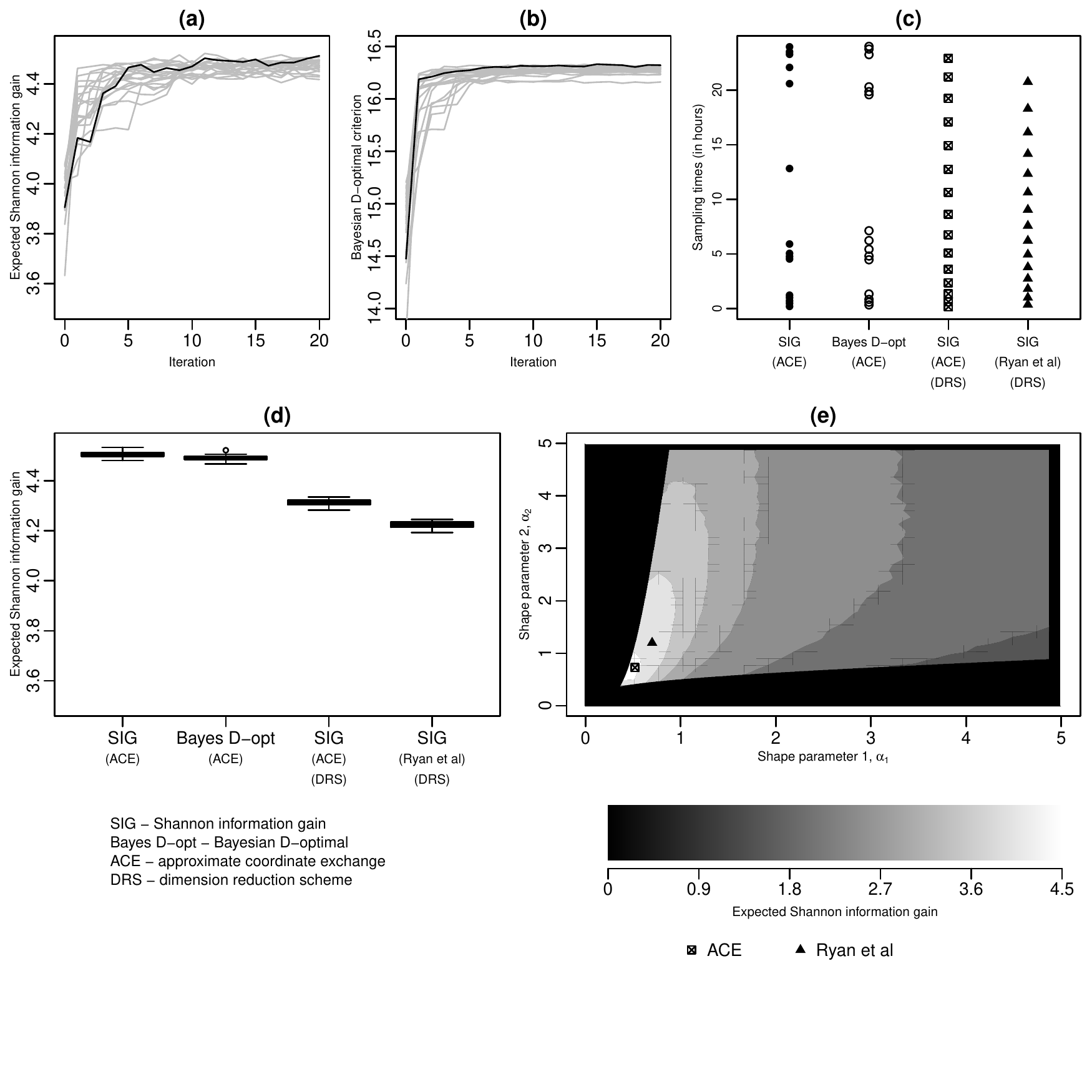}
\vspace{-2cm}
\caption{\label{PK_PIC1}(a), (b) Trace plots of $\tilde{U}(\boldsymbol{\delta}^C)$ for each iteration of the ACE algorithm for SIG and pseudo-Bayesian $D$-optimality utilities, respectively; in each plot, the black line shows the trace of the expected utility for the best design; (c) Designs found from the ACE algorithm: unrestricted SIG-optimal, pseudo-Bayesian $D$-optimal, Beta DRS SIG-optimal, together with the \citet{R2014} Beta DRS SIG-optimal designs; (d) Boxplots for 20 evaluations of $\tilde{U}^S(\bdelta^\star)$ for designs from these four methodologies; (e) Approximate expected utility surface for SIG as a function of the Beta DRS parameters; parameter values corresponding to the \citet{R2014} and the ACE DRS designs are marked.}
\end{figure}

For this problem, \citet{R2014} assumed that
$$
a(\btheta) = \frac{400\theta_2}{\theta_3(\theta_2 - \theta_1)}\,,\quad b(\btheta;t_i) = \left( 1 + \frac{a(\btheta)^2\mu(\btheta;t_i)^2}{10} \right)\,,\quad \sigma^2=0.1\,,
$$
and found designs using the SIG utility function. Independent log-normal prior distributions were assumed for the elements of $\boldsymbol{\theta}$ with, on the log scale, each having common variance 0.05 and expectations $\log(0.1)$, $\log(1)$ and $\log(20)$ for $\theta_1$, $\theta_2$ and $\theta_3$, respectively. These authors also incorporated the constraint $\max_{s,t=1,\dots,n}|t_s - t_t| \ge 0.25$, i.e. that sampling times must be at least 15 minutes apart. It is straightforward to incorporate this constraint into design search using the ACE algorithm. In Step~\ref{acceptstep1}, $\hat{U}(\delta_i\given\bdelta_{(i)}^C)$ is maximized over a set $\mD_i$ that satisfies the constraint. Phase II of the ACE algorithm is then omitted as replicated sampling times are not permitted.

\citet{R2014} restricted their search for a SIG-optimal design to the class of designs defined via a dimension reduction scheme (DRS) that set the $n$ sampling times to scaled percentiles of a Beta$(\alpha_1,\alpha_2)$ distribution. Hence, the design problem was reduced to selecting two parameters, $\alpha_1$ and $\alpha_2$. The \citet{M1999} simulation algorithm was used to sample from an artificial posterior distribution for $\alpha_1,\alpha_2$, with unnormalized density equal to the integrand in~\eqref{eq:exputil}. The chosen design was then the scaled quantiles from the Beta distribution obtained from using the posterior modal values of $\alpha_1$ and $\alpha_2$.  

We compare this design with three designs found from ACE: (i) a SIG-optimal design; (ii) a pseudo-Bayesian $D$-optimal design ; and (iii) an optimal choice of $\alpha_1,\alpha_2$ for the Beta DRS. For this latter design, the sampling times are given by $t_j = 24 \times Q\left(\frac{j}{n+1};\,\alpha_1,\alpha_2\right)$, with $Q(r;\,\alpha_1,\alpha_2)$ the $r$th quantile of the $\mathrm{Beta}(\alpha_1,\alpha_2)$ distribution. In Step~\ref{acceptstep1} of the ACE algorithm, the sets $\mathcal{D}_1$ and $\mathcal{D}_2$ are given by
\begin{eqnarray*}
\mathcal{D}_1 & = & \left\{ x \in \mathbb{R}^+ : \min_{j=1,\dots,n-1} \left| Q\left(\frac{j}{n+1},x,\alpha_2\right) - Q\left(\frac{j+1}{n+1},x,\alpha_2\right) \right|>\frac{0.25}{24}\right\}\,,\\
\mathcal{D}_2 & = & \left\{ x \in \mathbb{R}^+ : \min_{j=1,\dots,n-1} \left| Q\left(\frac{j}{n+1},\alpha_1,x\right) - Q\left(\frac{j+1}{n+1},\alpha_1,x\right) \right|>\frac{0.25}{24}\right\}\,.
\end{eqnarray*}

For the SIG- and $D$-optimal designs, Figures~\ref{PK_PIC1}(a) and~\ref{PK_PIC1}(b) show trace plots of the approximate expected utility at each iteration of the algorithm. Approximate convergence is demonstrated, for both utility functions, through each run of the algorithm resulting in a similar value of $\tilde{U}(\bdelta)$ after a relatively small number of iterations. Convergence is, however, achieved more quickly for pseudo-Bayesian $D$-optimality, which does not require approximation of posterior quantities. This criterion also displays greater consistency in the final approximated expected utility between runs of the algorithm.

The sampling times for the four designs, shown in Figure~\ref{PK_PIC1}(c), indicate that the designs using dimension-reduction do not display the clustering of points evident in the SIG and pseudo-Bayesian $D$-optimal designs. The boxplots in Figure~\ref{PK_PIC1}(d), from 20 evaluations of $\tilde{U}^S(\bdelta^\star)$ ($B=20,000$) for each design, confirm that larger approximate expected utilities are obtained, up to a 5\% improvement, when DRS is not used. Here, the pesudo-Bayesian $D$-optimal design provides a good approximation to the SIG-optimal design. 


The DRS design found from ACE outperforms the \citet{R2014} design. To explore this result further, the expected utility surface was investigated as a function of $\alpha_1$ and $\alpha_2$ by sampling 40,000 $(\alpha_1,\alpha_2)$ pairs from $[0,5]^2$ and evaluating $\tilde{U}^S(\bdelta)$ for each pair. The resulting expected utility surface is shown in Figure~\ref{PK_PIC1}(e), where $\tilde{U}^S(\bdelta)=0$ for parameter pairs that do not satisfy the 15 minute constraint. Both methods identify the relatively small region of high expected utility; the sampling-based algorithm (\citealp{R2014}, \citealp{M1999}) fails to identify the optimum point within this region.

\subsection[Logistic regression]{Logistic regression in four factors}
\label{LOGREG_SEC}
Fully-Bayesian design for multi-variable logistic regression has not appeared in the literature, although \citet{H2001} found a SIG-optimal design for a single-variable model and \citet{W2006} were the first to find multi-variable pseudo-Bayesian $D$-optimal designs. Here, we find designs for a first-order logistic regression model in four variables where the response is measured for $G$ groups of $n_g$ runs, i.e., $n=Gn_g$. Let $y_{st}\sim \mathrm{Bernoulli}(\rho_{st})$ be the $t$th response from the $s$th group ($s=1,\dots,G;\,t=1,\dots,n_g$), with
\begin{eqnarray*}
\log \left( \frac{\rho_{st}}{1-\rho_{st}}\right) & = & \beta_0 + \omega_{s0} + (\beta_1+\omega_{s1}) x_{1st} + (\beta_2+\omega_{s2}) x_{2st} + (\beta_3+\omega_{s3}) x_{3st}\\
& &\qquad +\, (\beta_4+\omega_{s4}) x_{4st}\nonumber \\
& = & \mathbf{x}_{st}^\T \left(\boldsymbol{\beta}+\boldsymbol{\omega}_s\right)\,,
\label{LOGMOD}
\end{eqnarray*}
where $\boldsymbol{\beta} \in \mathbb{R}^5$ are the parameters of interest, and $\boldsymbol{\omega}_s \in \mathbb{R}^5$ ($i=s,\dots,G$) are the group-specific nuisance parameters (or \textquotedblleft random effects\textquotedblright). Let $\bX = \left(\bX_1^\T\,\cdots\,\bX_G^\T\right)^\T$ be the $n \times 5$ model matrix where $\bX_s$ is the $n_g \times 5$ matrix with $t$th row given by $\mathbf{x}_{st}^\T$. The design matrix $\mathbf{D}$ is formed as the last four columns of $\mathbf{X}$, $\bdelta=\mathrm{vec}(\bD)$ has length $q=4n$, and $\mD_i=[-1,1]$ for $i=1,\ldots,q$.

The following independent prior distributions for each element of $\boldsymbol{\beta}$ are assumed:
\begin{equation}\label{eq:logisticprior}
\begin{array}{cccccccc}
\beta_0 \sim \mathrm{U}[-3,3]\,, & & \beta_1\sim\mathrm{U}[4,10]\,, & &\beta_2\sim\mathrm{U}[5,11]\,,\\ 
\beta_3\sim \mathrm{U}[-6,0]\,, & & \beta_4 \sim \mathrm{U}[-2.5,3.5]\,.\\ 
\end{array}
\end{equation}
We find designs for two different prior distributions for each $\bomega_s$ $(s=1,\dots,G)$: (i) a prior point mass at $\boldsymbol{\omega}_s=\mathbf{0}$ for all $s$, resulting in standard logistic regression with homogeneous groups; (ii) a hierarchical prior distribution in which the elements of $\boldsymbol{\omega}_s$ are independent and identically distributed as $\omega_{sr} \sim \mathrm{U}[-\lambda_r,\lambda_r]$,
for $r=0,\dots,4$, with $\lambda_r>0$ unknown and having triangular prior density $\pi(\lambda_r) = \frac{2(L_r - \lambda_r)}{L_r^2}$ with $(L_0,\dots,L_4) = (3,3,3,1,1)$.

\subsubsection{Logistic regression with homogeneous groups}\label{lsecloghom}

\begin{figure}
\centering
\includegraphics[width=154mm]{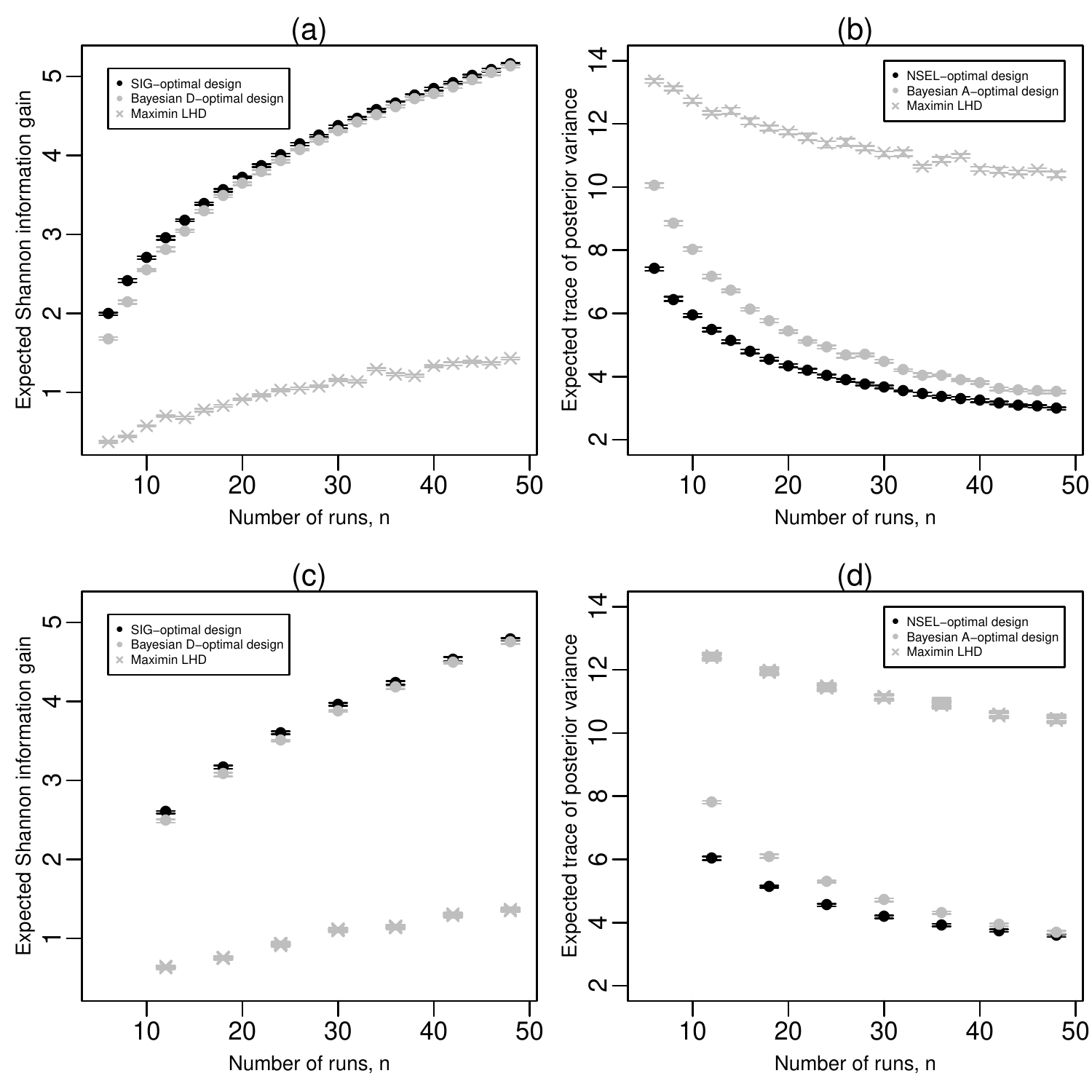}
\caption{\label{LOGREG_PIC}Results from 20 evaluations of (a) $\tilde{U}^S(\bdelta)$ for SIG-optimal, pseudo-Bayesian $D$-optimal and maximin Latin hypercube designs, and (b) $-\tilde{U}^V(\bdelta)$ for NSEL-optimal, pseudo-Bayesian A-optimal and maxmin Latin hypercube designs, for homogenous logistic regression; (c) and (d) show the same evaluations for hierarchical logistic regression. For the latter two plots, for each value of $n$, 20 different random assignments are made of the points of the Latin hypercube design to the $G$ groups, and each resulting design is evaluated 20 times. For each design, the central plotting symbol denotes the mean expected Shannon information gain or expected average posterior variance, with the two horizontal lines denoting the minimum and maximum of these quantities.}
\end{figure}

We use ACE to find designs that maximize the SIG and NSEL expected utilities for homogeneous logistic regression with $\bomega_s=0$ and $n=6,\ldots,48$. For comparison, we also find pseudo-Bayesian $D$- and $A$-optimal designs. We also compare to maximin Latin hypercube (LH) designs \citep{MM1995}. For this example, the starting designs for the algorithm were a locally $D$-optimal design (for SIG and Bayesian $D$) and a locally $A$-optimal design (for NSEL and Bayesian $A$), found from ACE via maximization of $\psi^S(\bdelta)$ or $\psi^V(\bdelta)$, respectively, using a point prior distribution for each parameter with support at the mean of each prior distribution in~\eqref{eq:logisticprior}. Figure~\ref{LOGREG_PIC} presents results (minimum, mean, maximum) of 20 evaluations of (a) $\tilde{U}^S(\bdelta)$ for the SIG-optimal, Bayesian $D$-optimal and maximin LH designs, and (b) $-\tilde{U}^V(\bdelta)$ for the NSEL-optimal, Bayesian $A$-optimal and maximin LH designs, using $B=20,000$ Monte Carlo samples. For small $n$, on average there are substantial differences in expected utility between the fully-Bayesian and pseudo-Bayesian designs, with the SIG-optimal design having expected Shannon information gain up to 20\% larger than the Bayesian $D$-optimal design and the NSEL-optimal design having expected trace of the posterior covariance matrix up to 27\% smaller than the Bayesian $A$-optimal design. For both SIG and NSEL, as $n$ increases, the difference in expected utility between these designs and the pseudo-Bayesian designs decreases. For SIG, these findings agree with asymptotic results on the convergence, under certain regularity conditions, of the posterior distribution to a normal distribution \citep[see, for example,][pp. 585-588]{G2014}. The maximin LH designs, which are model-free space-filling designs, perform poorly under both SIG and NSEL utilities and are not competitive with the model-based designs.  

As there are no comparable results on fully-Bayesian design for multi-variable logistic regression in the literature, we compare the pseudo-Bayesian $D$-optimal designs for $n=16$ and $n=48$ found from ACE with designs from the approach of \citet{GJS2009}. We independently implemented the methodology of these authors to obtain designs for $n=16$ and $n=48$; we also compare to the $n=16$ run design published by \citet{GJS2009}. For each of these three designs, we calculated the average of $D$-efficiency~\eqref{eq:Deff} over 20 Monte Carlo approximations (each with $B=20,000$) relative to the appropriately-sized design from ACE. The published 16-run design has average efficiency of 82\%; the designs from our implementation perform similarly to the ACE designs, with average efficiencies of $99.9\%$ and $101.3\%$ for $n=16$ and $n=48$, respectively. 

\subsubsection{Hierarchical Logistic Regression}

For hierarchical logistic regression, we again find SIG-optimal and NSEL-optimal designs, along with pseudo-Bayesian $D$- and $A$-optimal designs using an approximation to the Fisher information \citep[][p. 467]{pawitan}. We set $n_g=6$ and $G=2,\ldots,8$, leading to $n=12, 14, \ldots, 48$. To reduce the computational burden, $B=1000$ was used in Step~\ref{acceptstep2} to find SIG-optimal designs. Previous research has found pseudo-Bayesian $D$-optimal designs for smaller numbers of variables and group sizes \citep{WW2015}.

Figures~\ref{LOGREG_PIC}(c) and (d) show results from 20 evaluations of $\tilde{U}^S(\boldsymbol{\delta})$ and $-\tilde{U}^V(\boldsymbol{\delta})$ for the SIG-optimal and pseudo-Bayesian $D$-optimal designs, and the NSEL-optimal and pesudo-Bayesian $A$-optimal designs, respectively. Again, the performances of maximin LH designs are included for reference (see figure caption for details).  A comparison with Figures~\ref{LOGREG_PIC}(a) and (b), respectively, shows lower expected gains in Shannon information and higher expected posterior variance for the hierarchical logistic regression model due to additional uncertainty introduced by the group-specific parameters. As with designs for homogeneous logistic regression, the difference in expected utility between the pseudo-Bayesian designs and the fully-Bayesian designs decreases as $n$ increases, and the LH designs perform poorly.

\subsection{Binomial regression under model uncertainty}


Uncertainty over the choice of statistical model $\pi(\by,\bpsi\given\bdelta)$ is common in practice, and has been addressed in pseudo-Bayesian design for generalized linear models by \citet{W2006}. To demonstrate Bayesian optimal design under model uncertainty, we find follow-up designs for the beetle mortality study of \citet{B1935}, a common example used to illustrate binomial regression. In the original data set, 481 beetles were each administered one of eight different doses (in mg/L) of carbon disulphate.  We broadly follow the case study analysis of \citet[pp. 423-433]{OF2004}, who reproduced the data, and assume interest lies in providing a model-averaged posterior distribution of the  lethal dose 50 (LD50), the dose required to achieve 50\% mortality. 

We assume that the binary indicator of death for each beetle is an independent Bernoulli random variable. The number, $y_k$, of deaths from dose $x_k$ is modelled as $y_k \sim \mathrm{Binomial}(n_k,\rho_k)$, where $\rho_k$ is the probability of death for the $k$th dose which was administered to $n_k$ beetles, $\sum_{k=1}^nn_k = 481$. We denote the link function by $g(\rho_k) = \eta_k$, with $\eta_k$ the linear predictor and consider six models formed by the Cartesian product of three link functions and two linear predictors: the logit, $g(\rho_k) = \log \left\{\rho_k/(1-\rho_k)\right\}$, the c-log-log, $g(\rho_k) = \log \left\{ - \log \left( 1 - \rho_k\right)\right\}$, and the probit, $g(\rho_k) = \Phi^{-1}\{\rho_k\}$, with $\Phi\{\cdot\}$ the standard normal distribution function; and 1st order ($\eta_i = \beta_0 + \beta_1 x_i$) and 2nd order ($\eta_i = \beta_0 + \beta_1 x_i + \beta_2 x_i^2$) linear predictors. 

Let $u \in \mathcal{U} = \left\{1,\dots,6 \right\}$ denote the model indicator (see Table~\ref{BEETLE2}) and let $\boldsymbol{\beta}_u$ denote the vector of regression parameters under model $u$. LD50 is then  given by
$$LD(\bbeta_u) = \left\{ \begin{array}{ll}
\frac{w - \beta_{0}}{\beta_{1}} & \mbox{for $u = 1,3,5$ (1st order linear predictor)}\\
\frac{-\beta_{1} + \sqrt{\beta_{1}^2 - 4 \beta_{2} \left(\beta_{0} - w\right)}}{2\beta_{2}} & \mbox{otherwise}\,,
\end{array} \right.$$
where $w = \log \left\{-\log \left(0.5\right)\right\}$ for the c-log-log link function, and $0$ otherwise. We use unit information prior distributions \citep{N2003} for $\boldsymbol{\beta}_u\given u$ under each model and set $\pi(u)=1/6$ for $u=1,\ldots,6$. The posterior model probabilities for each model are approximated using importance sampling to evaluate the marginal likelihood of each model, and given in Table~\ref{BEETLE2}. Samples from the posterior distribution of the model parameters are generated for each of the six models using the Metropolis-Hastings algorithm, and then weighted by $\pi(u\given\by)$ to produce a sample from the joint posterior distribution $\boldsymbol{\beta}_u,u\given\mathbf{y}$ of regression parameters and model indicator. A sample from the model-averaged posterior distribution of LD50 can be obtained by evaluating $LD(\bbeta_u)$ for each sampled parameter vector.

\begin{table}[ht]
\caption{Approximate posterior model probabilities, $\pi(u\given\by)$, for the beetle mortality data.\vspace{-0.2cm}}
\label{BEETLE2}
\begin{center}
\begin{tabular}{lllc} \hline
u & Link Function & Linear Predictor & $\pi(u\given\mathbf{y})$ \\ \hline
1 & Logit & 1st order & 0.0216 \\
2 & Logit & 2nd order & 0.0686 \\
3 & C-log-log & 1st order & 0.7580 \\
4 & C-log-log & 2nd order & 0.0612 \\
5 & Probit & 1st order & 0.0304 \\
6 & Probit & 2nd order & 0.0602 \\ \hline
\end{tabular}
\end{center}
\end{table}

We consider the design of a follow-up experiment using a further $n_0$ (potentially new) doses. Each dose is to be administered to $n_{0k_0}$ beetles ($k_0=1,\ldots,n_0$) and, in each group, the number, $y_{0k_0}$, of beetles that die is recorded. Let $\mathbf{y}_0$ be the $n_0 \times 1$ vector of the numbers of beetles that die in the follow-up experiment. We assume that $n_{0k_0}$ is unknown and adopt a Poisson$(\lambda)$ prior distribution. Hence $y_{0k_0} \sim \mathrm{Poisson}(\lambda\rho_{k_0})$. We choose $\lambda=60$, consistent with the values of $n_k$ in the original dataset, and find designs for $n_0 = 1,\ldots,10$ to estimate the value of LD50 under the NSEL utility function by maximizing
$$U^V(\bdelta) = - \sum_{u=1}^6\pi(u\given\by)\int_{\mathcal{Y}}  \int_{\mathcal{B}_u} \left[LD(\boldsymbol{\beta}_u) - \mathrm{E} \left( LD(\boldsymbol{\beta}_u)\given\mathbf{y}_0,\mathbf{y},\boldsymbol{\delta}\right)\right]^2\pi(\beta_u, \by_0\given u,\by)\mathrm{d}\bbeta_u\mathrm{d}\by_0\,,$$
where design $\boldsymbol{\delta}$ is the $n_0 \times 1$ vector of doses and $\mathcal{B}_u$ is the parameter space for model $u$. For the purposes of design and modelling, we assume that $\delta_i\in\mD_i=[-1,1]$ for all $i=1,\ldots,n_0$ but transform the doses to the original scale $[1.6907,1.8839]$ for the presentation of results.

\begin{figure}
\centering
\includegraphics[width=160mm]{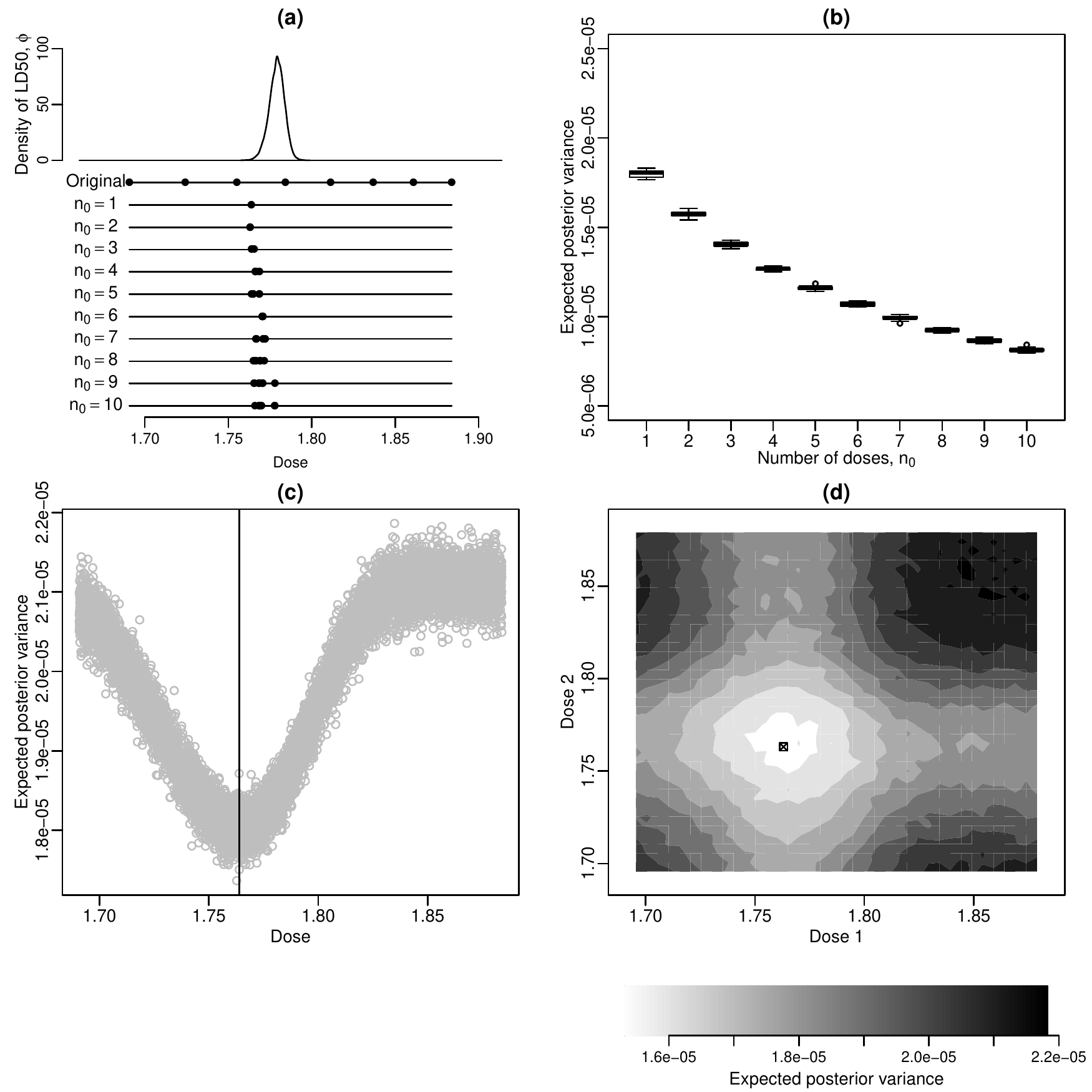}
\caption{\label{BEETLE_PIC}(a) Posterior density for LD50, the original experimental doses and optimal doses (in mg/L) for each value of $n_0$; (b) boxplots of 20 evaluations of $-\tilde{U}^V(\bdelta^\star)$ for each $n_0$ for the NSEL-optimal designs; (c) negative approximate expected utility $-\tilde{U}^V(\bdelta)$ against dose for $n_0=1$; the vertical line indicates $\bdelta^\star$. (d) negative approximate expected utility $-\tilde{U}^V(\bdelta)$ against dose for $n_0=2$; $\boxtimes$ indicates $\bdelta^\star$.}
\end{figure}

We can approximate $U^V(\boldsymbol{\delta})$ by
$$\tilde{U}^V(\boldsymbol{\delta}) = - \frac{1}{B} \sum_{l=1}^B \left[LD(\boldsymbol{\beta}_{ul}) - \hat{\E}\left(LD(\boldsymbol{\beta}_u)\given\mathbf{y}_{0l},\mathbf{y},\boldsymbol{\delta}\right)\right]^2\,,$$
where $\left\{\boldsymbol{\beta}_{ul},u_l,\mathbf{y}_{0l}\right\}_{l=1}^B$ is a random sample from the joint distribution with density $\pi(\boldsymbol{\beta}_u,u,\mathbf{y}_0\given\mathbf{y})$, and
$$\hat{\mathrm{E}}\left(LD(\boldsymbol{\beta}_m)\given\mathbf{y}_0,\mathbf{y},\boldsymbol{\delta}\right) = \frac{\sum_{b=1}^{\tilde{B}} 
LD(\tilde{\boldsymbol{\beta}}_{\tilde{u}b})\pi(\mathbf{y}_0\given\tilde{\boldsymbol{\beta}}_{\tilde{u}b},\tilde{m}_b)}{\sum_{b=1}^{\tilde{B}} \pi(\mathbf{y}_0\given\tilde{\boldsymbol{\beta}}_{\tilde{u}b},\tilde{m}_b)}\,,$$
where $\left\{\tilde{\boldsymbol{\beta}}_{\tilde{u}b},\tilde{m}_b\right\}_{b=1}^{\tilde{B}}$ is a random sample generated from the joint distribution with density $\pi(\boldsymbol{\beta}_u,u\given\mathbf{y})$.

Figure~\ref{BEETLE_PIC} summarises the results from the ACE algorithm. The doses in the NSEL-optimal design lie in the lower tail of the (original) posterior distribution of LD50 for all values of $n_0$, see Figure~\ref{BEETLE_PIC}(a). For $n_0>1$, the doses are concentrated near a single point ($1.77$), for example four replicate points occur for $n_0=10$. The approximate expected posterior variance of LD50, $-\tilde{U}^V(\boldsymbol{\delta})$, rapidly decreases as $n_0$ is initially increased from 1, see Figure~\ref{BEETLE_PIC}(b); the rate of decrease slows as $n_0$ becomes larger.

To further investigate the selected designs, the expected utility surface and the performance of the ACE algorithm, we randomly generated 10,000 designs for $n_0=1$ and $n_0=2$ uniformly from $\mD_1$ and $\mD_1^2$, respectively. For each design, we evaluate $-\tilde{U}^V(\boldsymbol{\delta})$ and plot against dose; see Figure~\ref{BEETLE_PIC}(c) for $n_0=1$ and Figure~\ref{BEETLE_PIC}(d) for $n_0=2$. The NSEL design identified by ACE is marked in each plot and, for both values of $n_0$, the minimum negative expected utility is achieved. The variance of the original model-averaged posterior distribution for LD50 is $2.10 \times 10^{-5}$. Hence for both $n_0=1$ and $n_0=2$, it is clear that choosing a design composed of only very high or low doses would have resulted in a negligible expected reduction in variance.

\section[Discussion]{Discussion and future work}\label{DISC}
The ACE methodology proposed in this paper provides a step-change in the nature and complexity of statistical models and experiments for which Bayesian designs can be obtained. It may be used to find decision-theoretic designs whenever it is possible to simulate values from the prior distribution of the model parameters and responses from the statistical model. The combination of emulating an approximation to the expected utility and the coordinate exchange algorithm has allowed much larger problems to be tackled than was previously possible, both greater numbers of runs and more controllable variables. The algorithm also matches or exceeds the performance of existing approaches for smaller problems, and offers a clear advantage for design selection over the application of a dimension reduction scheme. The new designs made possible by this methodology also allow previously impossible benchmarking of designs from asymptotic approximations.

As presented, ACE can be applied to numerous important practical problems using the available \texttt{R} package. We have applied, or are in the process of applying, ACE to problems from chemical development and biological science. There are also a variety of extensions that could be made to ACE to increase its computational efficiency and applicability. We now highlight a few of these areas.

In ongoing work, we are extending and applying the methodology to find designs for statistical models where the likelihood function is only available numerically as the output from an expensive computer code (see also \citealp{HM2013}). Such models include those described by the solution to a system of non-linear differential equations, which are increasingly studied in the field of uncertainty quantification (e.g. \citealp{CCCG2015}).

Convergence of the algorithm may be improved through a reparameterization of the design to remove dependencies between coordinates (e.g. \citealp{Fletcher1987}, p. 19) that can be evident in efficient designs for some models. Such dependencies could be identified through pilot runs of the algorithm or by studying properties of pseudo-Bayesian designs. Additionally, the computational burden of the algorithm could be further reduced by employing alternative approaches to perform each one-dimensional optimization step in the algorithm. For example, a sequential strategy could use an expected improvement criterion modified for stochastic responses (e.g. \citealp{PGRC2013}).

Alternative strategies could also be adopted for the approximation of the expected utility. Zero-variance Monte Carlo (\citealp{Ripley1987}, pp. 129-132, \citealp{msi2013}) could be used to reduce the variance of the Monte Carlo estimator through the introduction of negative correlations via antithetic variables. Combining deterministic approximations, such as expectation propagation, with Monte Carlo methods would remove the need for nested simulation and may work well for nonlinear regression models with normal prior distributions.



\section*{Acknowledgements}
This work was supported by the U.K. Engineering and Physical Sciences Research Council through Fellowship EP/J018317/1 for D.C. Woods. The authors thank the participants at the ``Bayesian Optimal Design of Experiments'' workshop (Brisbane, Australia, December 2015; \texttt{http://www.bode2015.wordpress.com}) for useful discussions on extensions and future work. The authors acknowledge the use of the IRIDIS High Performance Computing Facility, and associated support services at the University of Southampton, in the completion of this work.

\bibliographystyle{asa}
\bibliography{mybib}
\end{document}